\documentclass[%
 reprint,
 amsmath,amssymb,
 aps,
]{revtex4-2}
\usepackage{mathptmx}
\usepackage{etoolbox}
\usepackage{xcolor}
\usepackage{hyperref}
\usepackage{graphicx}
\usepackage{dcolumn}
\usepackage{bm}

\begin{document}

\preprint{APS/123-QED}

\title{Charge Density Wave Order and Superconductivity in Janus MoXH Monolayers}

\author{Jakkapat Seeyangnok$^{1}$}
 \email{jakkapatjtp@gmail.com} 
\author{Udomsilp Pinsook$^{1}$}%
 \email{Udomsilp.P@Chula.ac.th}

\author{Graeme J Ackland$^{2}$}
 \email{gjackland@ed.ac.uk} 
\affiliation{$^{1}$Department of Physics, Faculty of Science, Chulalongkorn University, Bangkok, Thailand.\\
$^{2}$Centre for Science at Extreme Conditions, School of Physics and Astronomy, University of Edinburgh, Edinburgh, United Kingdom}%


\date{\today}

\begin{abstract}
Two-dimensional Janus hydrogenated transition-metal chalcogenides provide an unusual platform where lattice instabilities, electron--phonon coupling, and superconductivity are strongly intertwined. Using first-principles calculations, we demonstrate that Janus 2H/1T MoXH ($X=\mathrm{S,Se}$) monolayers host an intrinsic, commensurate charge-density-wave (CDW) ground state originating from soft phonon modes at the Brillouin-zone $M$ point. Real-space supercell optimizations confirm that the CDW reconstruction lowers the total energy and fully stabilizes the lattice, eliminating the imaginary phonon modes present in the high-symmetry metallic structures. Analysis of the electronic susceptibility shows that the CDW instability is not driven by Fermi-surface nesting, but instead arises from strong electron--phonon coupling. We further reveal a material-dependent interplay between CDW order and superconductivity. In 1T--MoSH, CDW formation enhances low-energy phonon contributions and strengthens electron--phonon coupling, leading to an increased superconducting transition temperature. In contrast, for 1T--MoSeH and 2H--MoSeH, the CDW phase suppresses electron--phonon coupling and reduces superconductivity. Finally, we show that thermal fluctuations, compressive strain, and carrier doping can selectively suppress CDW order and restore superconductivity. These results establish Janus MoXH monolayers as a tunable two-dimensional system for exploring lattice-driven charge ordering and its competition with superconductivity.
\end{abstract}

\maketitle


    Charge-density-wave (CDW) order is a ubiquitous collective phenomenon in low-dimensional solids, characterized by a periodic modulation of the electronic charge density accompanied by a symmetry-lowering lattice distortion~\cite{gruner2018density}. The concept of a CDW was originally formulated within the Peierls framework for one-dimensional metals, where perfect Fermi-surface nesting leads to a divergent electronic susceptibility and a lattice instability at the nesting wave vector~\cite{peierls1955quantum}. Subsequent developments established that the electronic instability is accompanied by a softening of phonon modes through electron--phonon coupling, a phenomenon later formalized as the Kohn anomaly~\cite{kohn1959image}. However, extensive theoretical and experimental investigations have demonstrated that this idealized Peierls scenario is rarely realized in real materials beyond strictly one-dimensional systems. In quasi-two-dimensional materials, Fermi-surface nesting alone is generally insufficient to drive CDW formation, and the transition instead emerges from a cooperative interplay between electronic states and lattice degrees of freedom, governed by a strongly momentum-dependent electron--phonon coupling~\cite{johannes2008fermi,zhu2017misconceptions}. Consequently, CDWs in realistic materials are now understood as coupled electronic--structural phase transitions rather than purely electronic instabilities, with phonon softening at specific wave vectors serving as a key microscopic signature of the transition~\cite{zhu2017misconceptions}.

    Within this framework, transition-metal dichalcogenides (TMDs) constitute a prototypical platform for realizing CDW order in two dimensions. Layered TMDs such as NbSe$_2$~\cite{ silva2016electronic,weber2011extended,ugeda2016characterization, nakata2021robust}, TaSe$_2$~\cite{johannes2008fermi, nakata2021robust, xi2015strongly}, and TaS$_2$~\cite{dalal2025flat,tsen2015structure, philip2023local} exhibit robust CDW phases despite the absence of perfect Fermi-surface nesting, indicating that their CDW transitions cannot be understood within a purely electronic Peierls picture~\cite{johannes2008fermi,zhu2017misconceptions}. Instead, the instability is driven by a strongly momentum-dependent electron--phonon coupling that selectively softens specific phonon modes and stabilizes a symmetry-lowering lattice distortion~\cite{weber2011extended,calandra2011charge}. Owing to weak interlayer coupling, this CDW order persists down to the monolayer limit, where genuine two-dimensional CDW phases have been experimentally observed in systems such as NbSe$_2$ and TaS$_2$~\cite{ugeda2016characterization,xi2015strongly}. These features establish TMDs as an ideal platform for exploring CDW physics beyond one dimension and its interplay with reduced dimensionality and competing electronic orders, including superconductivity.

    More recently, the successful synthesis of a Janus 2H--MoSH monolayer using the SEAR technique has stimulated renewed interest in hydrogenated Janus systems~\cite{lu2017janus}. Subsequent theoretical studies have predicted conventional phonon-mediated superconductivity in the 2H phase, with superconducting transition temperatures approaching 27~K and without evidence of a CDW instability in the high-symmetry structure~\cite{liu2022two}. In contrast, the corresponding 1T phase has been suggested to exhibit enhanced lattice instabilities associated with possible CDW formation~\cite{ku2023ab}. More recently, evidence of CDW ordering has also been reported in both the 2H and 1T phases of MoSeH~\cite{sui2025two}. Similar behavior has been proposed for related hydrogenated Janus systems such as WSH and WSeH, where superconductivity with $T_c$ values of order 10--15~K is accompanied by structural distortions associated with CDW ordering~\cite{seeyangnok2024superconductivity,seeyangnok2024superconductivitywseh,qiao2024prediction}. In addition, several Janus MXH monolayers ($M=\mathrm{Ti,Zr,Hf}$; $X=\mathrm{S,Se,Te}$) have been theoretically investigated and predicted to exhibit phonon-mediated superconductivity with $T_c$ values ranging from 9 to 30~K~\cite{li2024machine,ul2024superconductivity,seeyangnok2025competition}, where some of these systems also display lattice instabilities indicative of CDW phases similar to those in MoSH-, WSH-, and WSeH-based monolayers.
    
    Despite these advances, a systematic understanding of lattice instabilities and their interplay with superconductivity in hydrogenated Janus TMDs is still incomplete. In particular, soft phonon modes related to CDW phases are often observed in high-symmetry metallic structures and are usually ignored by stabilizing the system through external strain or carrier doping. However, the possibility that these instabilities lead to lower-symmetry CDW ground states has been largely overlooked. In this work, we address this issue by performing comprehensive first-principles calculations on Janus 2H/1T MoXH monolayers ($X=\mathrm{S,Se}$). We demonstrate that these systems host an unexpected, commensurate CDW ground state driven by soft-phonon instabilities at the Brillouin-zone $M$ point. We characterize the associated structural reconstruction, its energetic stabilization, and its impact on electron–phonon coupling and superconductivity. Furthermore, we show that external control parameters—including strain, carrier doping, and temperature—can suppress the CDW order and restore superconductivity. 
    
\begin{figure}[h!]
    \centering
    \includegraphics[width=8.5cm]{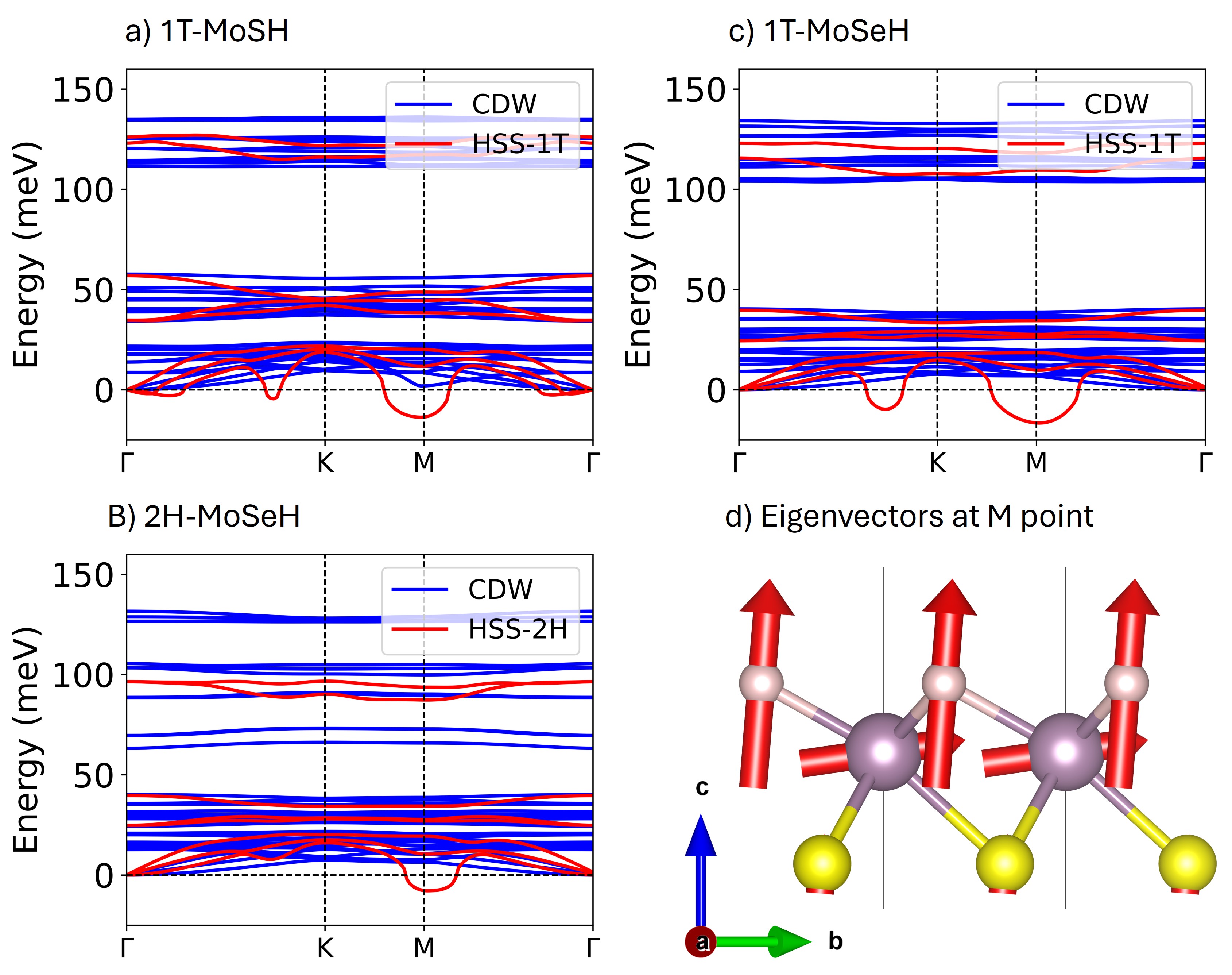}
    \caption{Phonon dispersions and lattice instabilities in hydrogenated Mo-based monolayers. \textbf{a--c}, Phonon dispersion relations of \textbf{(a)} 1T--MoSH, \textbf{(b)} 2H--MoSeH, and \textbf{(c)} 1T--MoSeH calculated for the high-symmetry structure (HSS, red lines) and the CDW phase (blue lines) along the high-symmetry path $\Gamma$--K--M--$\Gamma$.
    \textbf{d}, Atomic displacement patterns (phonon eigenvectors) of the soft mode at the $M$ point, illustrating the characteristic distortions dominated by the chalcogen and hydrogen atoms.}
    \label{fig:phonon_cdw_hss}
    \end{figure}

The high-symmetry structure (HSS) of Janus Mo-based chalcogenide (S, Se) hydrides belongs to the trigonal space group $P3m1$ (No.~156). We confirmed that the ground state of these materials favors a metallic non-magnetic configuration. The phonon dispersions calculated for the high-symmetry structure exhibit clear dynamical instabilities, as indicated by the red solid phonon branches in Fig.~\ref{fig:phonon_cdw_hss}(a--c). In particular, a soft vibrational mode with imaginary frequencies appears around the $M$ point, signaling that the high-symmetry metallic phase is unstable at zero temperature and tends toward a structural distortion.

Notably, the unstable phonon modes in all three systems exhibit similar eigenvectors, as shown in Fig.~\ref{fig:phonon_cdw_hss}~(d). These modes mainly involve collective atomic displacements of the Mo--chalcogen--H framework, indicating that the lattice lowers its energy by forming a distorted structure. Total-energy calculations confirm that the CDW phase is energetically favored over the high-symmetry structure for all three systems. The corresponding atomic displacements are consistent with a periodic lattice reconstruction associated with the emergence of a commensurate CDW phase, rendering the CDW-distorted structure more stable than the high-symmetry metallic configuration.


    \begin{figure}[h!]
    \centering
    \includegraphics[width=8cm]{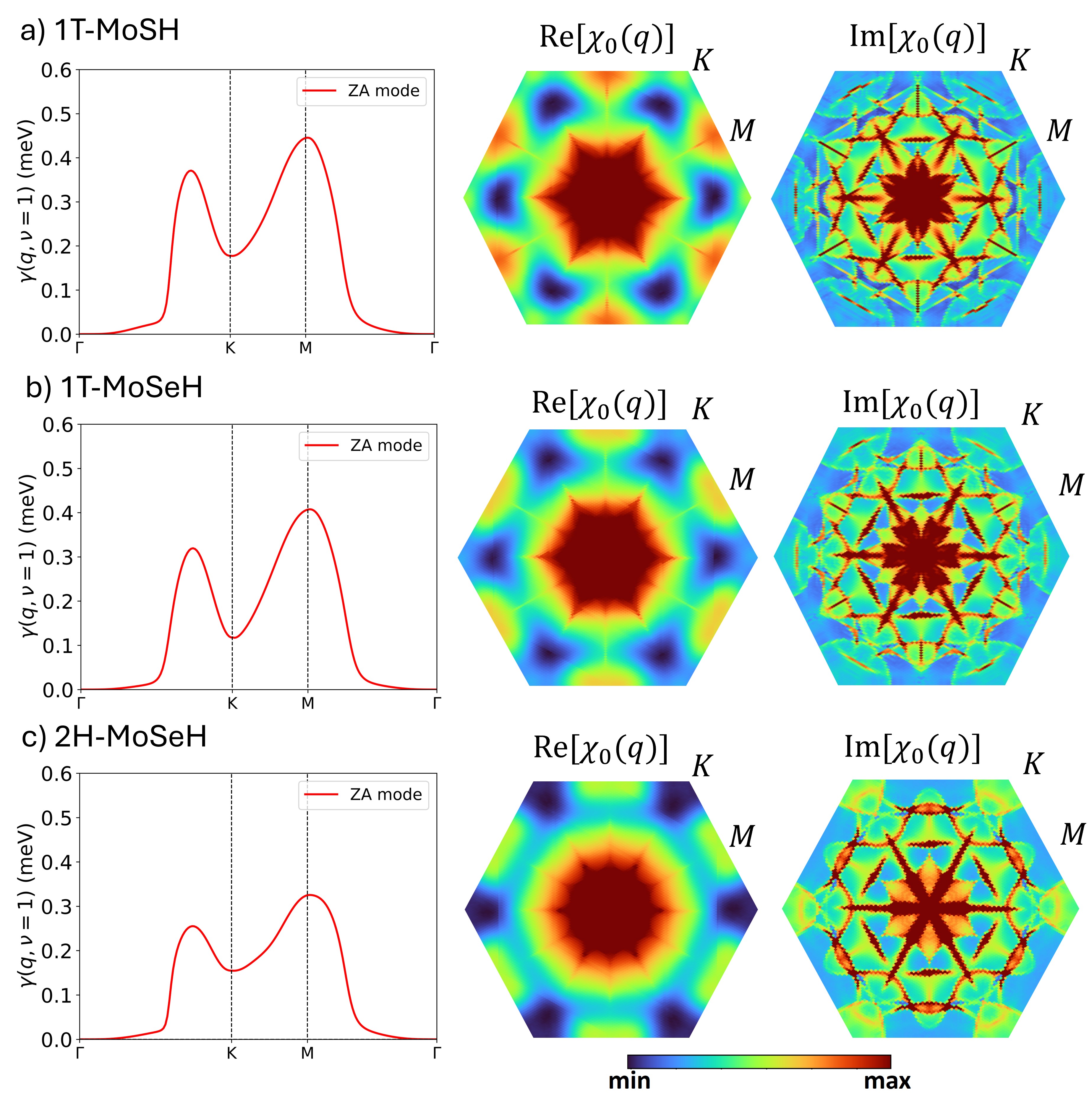}
    \caption{Interplay between phonon softening and electronic susceptibility in Janus MoXH monolayers.
    Rows correspond to
    (\textbf{a}) 1T--MoSH,
    (\textbf{b}) 1T--MoSeH, and
    (\textbf{c}) 2H--MoSeH.
    Left panels show the momentum-dependent phonon linewidth $\gamma(\mathbf{q},\nu{=}1)$ of the lowest-energy (ZA) phonon mode along the high-symmetry path $\Gamma$--K--M--$\Gamma$, exhibiting pronounced enhancement near the $M$ point.
    Middle panels display Brillouin-zone maps of the real part of the static electronic susceptibility $\mathrm{Re}[\chi_0(\mathbf{q})]$.
    Right panels show the corresponding imaginary part $\mathrm{Im}[\chi_0(\mathbf{q})]$.}
    \label{fig:phonon_susceptibility}
    \end{figure}

    \begin{figure*}
    \centering
	\includegraphics[width=18cm]{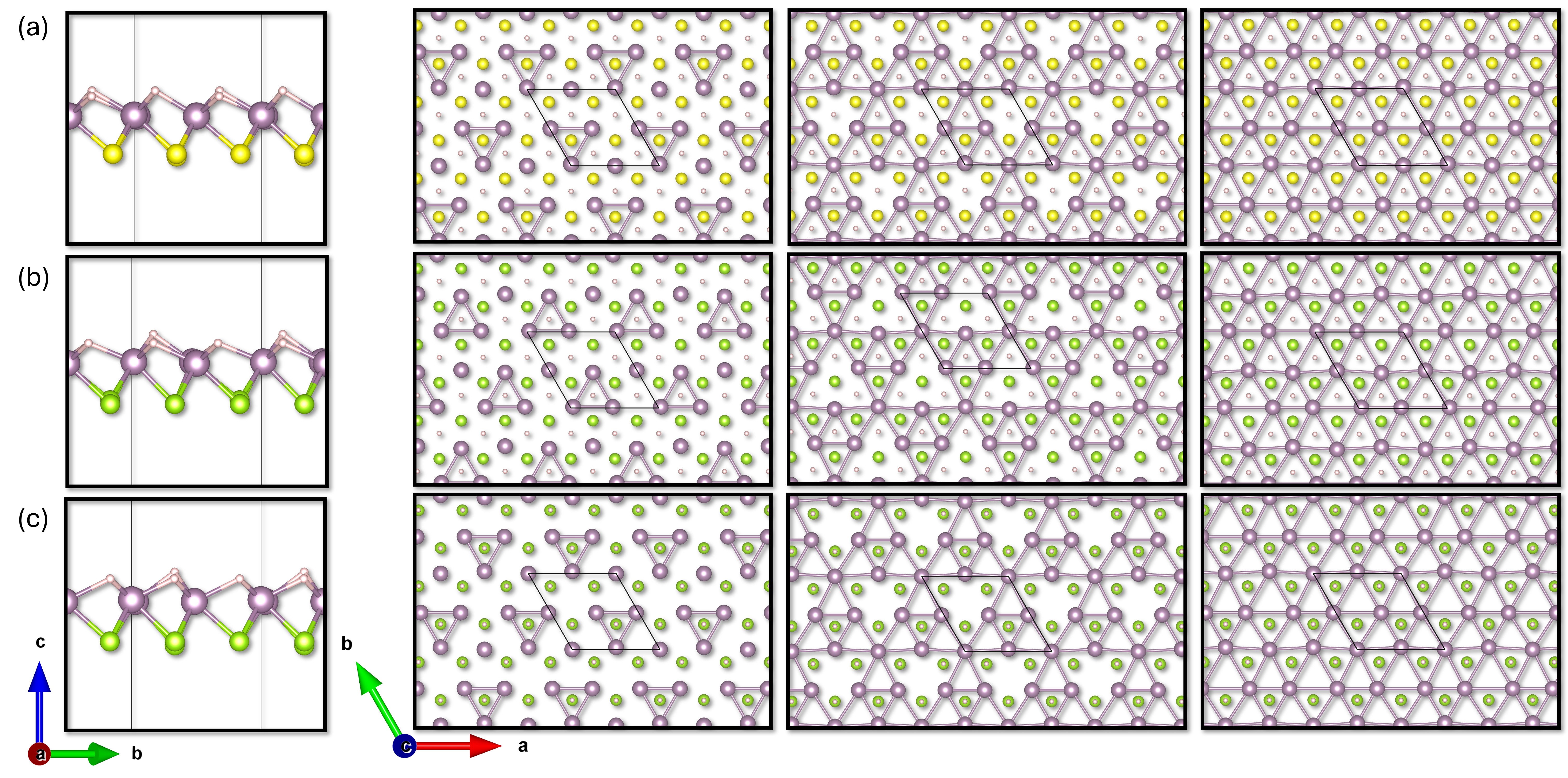}
	\caption{Real-space CDW structures in a $2\times2\times2$ supercell of Janus MoXH monolayers. Relaxed $2\times2$ CDW structures of (\textbf{a}) 1T--MoSH, (\textbf{b}) 1T--MoSeH, and (\textbf{c}) 2H--MoSeH.Side and top views illustrate the periodic lattice distortion associated with the commensurate CDW order.}
	\label{fig:cdw_realspace}
	\end{figure*}

To clarify the microscopic origin of the CDW instability in Janus Mo-based chalcogenide hydrides, we analyze the phonon dispersions, phonon linewidths, and the real and imaginary parts of the bare electronic susceptibility, as summarized in Fig.~\ref{fig:phonon_susceptibility}. In the static limit ($\omega \rightarrow 0$), the imaginary part of the susceptibility, Im$[\chi_0(\mathbf{q})]$, reflects the topology of the Fermi surface and is often associated with Fermi-surface nesting, whereas the real part, Re$[\chi_0(\mathbf{q})]$, directly governs the renormalization of phonon frequencies and the stability of the lattice.

For a CDW driven by an electronic instability, pronounced peaks in both Re$[\chi_0(\mathbf{q})]$ and Im$[\chi_0(\mathbf{q})]$ are expected to coincide at the CDW wave vector $\mathbf{q}_{\mathrm{CDW}}$. In the absence of such coincidence, the instability is instead dominated by momentum-dependent electron--phonon coupling. For the 1T--MoSH, 1T--MoSeH, and 2H--MoSeH phases, the phonon spectra exhibit pronounced softening of the out-of-plane acoustic mode near the $M$ point, indicating a lattice instability toward a commensurate modulation with $\mathbf{q}_{\mathrm{CDW}}\approx M$. Although the calculated Fermi surfaces display hexagonally warped contours with partial nesting features, the maxima of Im$[\chi_0(\mathbf{q})]$ are broadly distributed and do not uniquely coincide with the $M$ point. In contrast, Re$[\chi_0(\mathbf{q})]$ exhibits pronounced peaks near $M$, in direct correspondence with the momentum location of the phonon softening and the enhanced phonon linewidth shown in Fig.~\ref{fig:phonon_susceptibility}. The mismatch between the peak positions of Re$[\chi_0(\mathbf{q})]$ and Im$[\chi_0(\mathbf{q})]$ demonstrates that simple Fermi-surface nesting is not the primary driving force for the CDW transition~\cite{johannes2008fermi,zhu2017misconceptions}. Instead, the CDW instability is governed by strong momentum-dependent electron--phonon coupling.

Upon structural relaxation, the high-symmetry metallic phase spontaneously distorts into a lower-symmetry configuration characterized by periodic atomic displacements within a $2\times2$ in-plane supercell, as shown in Fig.~\ref{fig:cdw_realspace}. Structural optimizations using larger supercells confirm that this distortion is commensurate with the $M$-point wave vector. The resulting lattice reconstruction breaks the translational symmetry of the pristine $1\times1$ structure while preserving the overall hexagonal symmetry, consistent with a lattice-driven CDW transition.

The stabilization of this distorted phase confirms that the imaginary phonon modes observed at the $M$ point correspond to a genuine structural instability. Phonon dispersions calculated for the CDW-distorted structures are dynamically stable and show no imaginary frequencies throughout the Brillouin zone, in contrast to the high-symmetry phases. The removal of the soft modes upon structural distortion demonstrates that the CDW phase corresponds to the true vibrational ground state.

    \begin{figure}[h!]
    \centering
    \includegraphics[width=8cm]{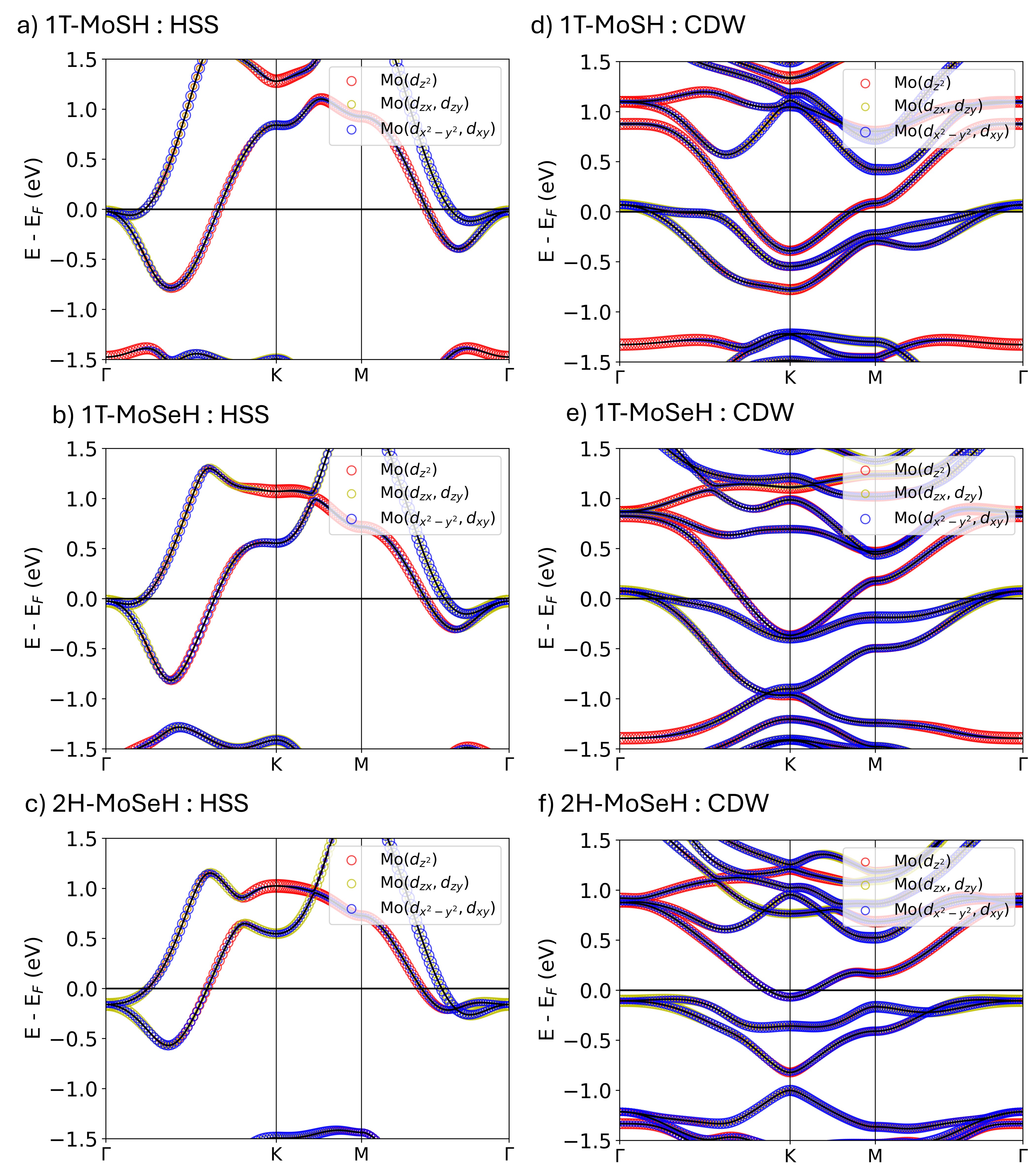}
    \caption{Orbital-resolved electronic band structures of Janus MoXH monolayers across the CDW transition. Orbital-resolved electronic band structures of
    (\textbf{a,d}) 1T--MoSH, (\textbf{b,e}) 1T--MoSeH, and
    (\textbf{c,f}) 2H--MoSeH, shown for the high-symmetry and CDW phases, respectively.}
    \label{fig:electronic_cdw}
    \end{figure}

The formation of the $2\times2$ CDW supercell leads to a pronounced reconstruction of the electronic bands due to Brillouin-zone folding and lattice distortion, as shown in Fig.~\ref{fig:electronic_cdw}. Despite this reconstruction, all CDW phases remain metallic, with multiple bands crossing the Fermi level. Orbital-resolved analysis reveals that the electronic states near the Fermi level are dominated by Mo $d$ orbitals, with substantial hybridization among the $d_{z^2}$, $d_{xz}/d_{yz}$, and $d_{x^2-y^2}/d_{xy}$ components.


    \begin{figure}[h!]
    \centering
    \includegraphics[width=8.5cm]{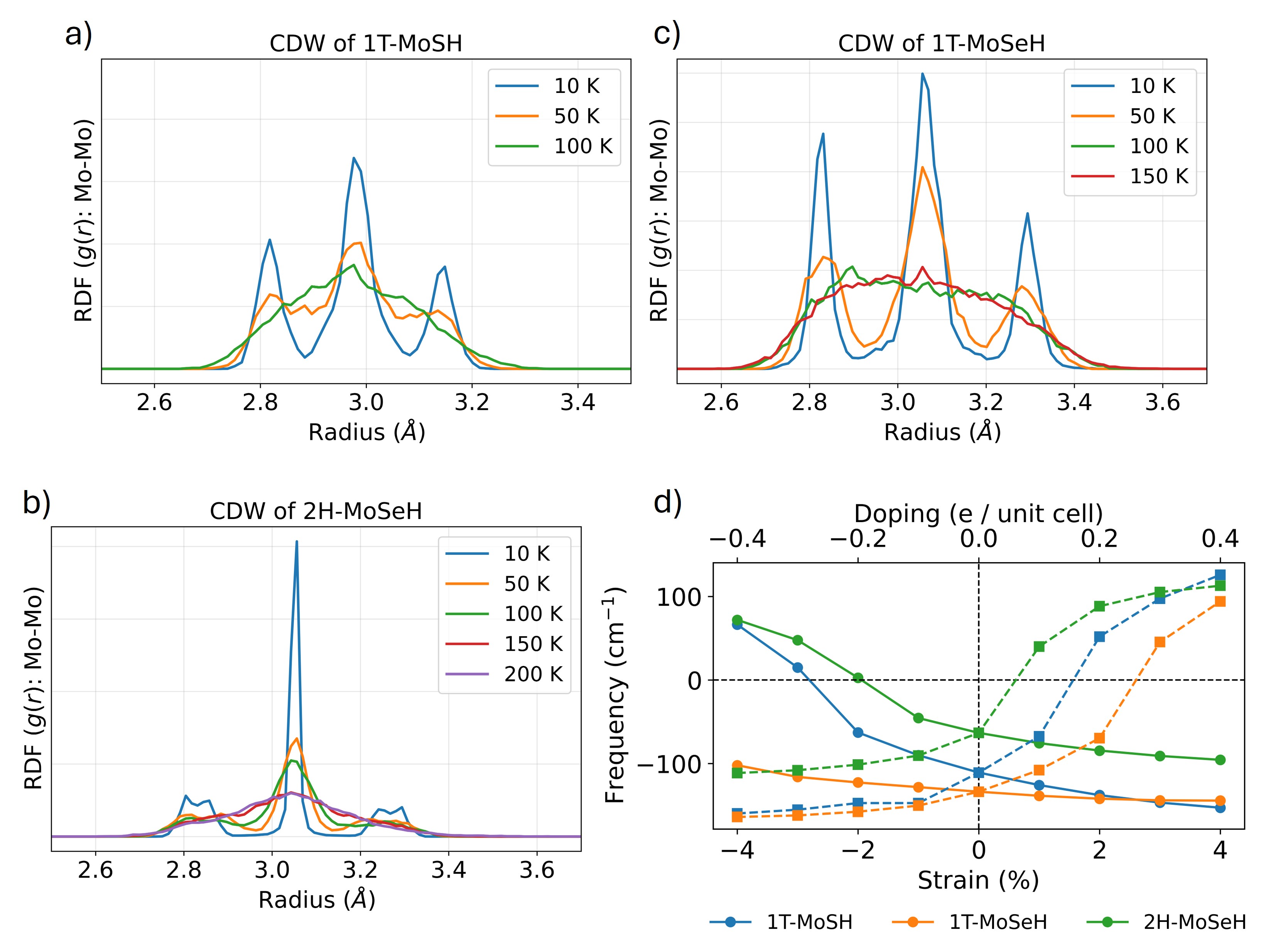}
    \caption{Thermal, strain, and carrier-doping control of CDW order in Janus MoXH monolayers.
    (\textbf{a--c}) Mo--Mo radial distribution functions (RDFs) of
    (\textbf{a}) 1T--MoSH,
    (\textbf{b}) 2H--MoSeH, and
    (\textbf{c}) 1T--MoSeH obtained from AIMD simulations at different temperatures.
    (\textbf{d}) Evolution of the CDW-related phonon frequency at the $M$ point as a function of biaxial strain and carrier doping.}
    \label{fig:cdw_control}
    \end{figure}

The stability of the CDW phase in Janus MoXH monolayers can be controlled by temperature, strain, and carrier doping, as summarized in Fig.~\ref{fig:cdw_control}. Analysis of the Mo--Mo radial distribution functions obtained from \textit{ab initio} molecular dynamics (AIMD) shows that, at low temperatures, the RDFs exhibit multiple well-separated peaks reflecting inequivalent Mo--Mo bond lengths induced by the CDW distortion. With increasing temperature, these peaks gradually merge into a single dominant peak, indicating thermal suppression of long-range CDW order.

The CDW instability in the high-symmetry structure can also be tuned by external strain and carrier doping. For 1T--MoSH and 2H--MoSeH, compressive biaxial strain hardens the soft phonon mode at the $M$ point and removes the imaginary frequency, thereby stabilizing the lattice against CDW formation. A similar stabilizing effect is achieved through hole doping, which suppresses the phonon softening associated with the CDW instability.

In contrast, the response of 1T--MoSeH is qualitatively different. Although moderate carrier doping initially stabilizes the soft phonon mode at the $M$ point, further doping induces a new lattice instability in the vicinity of the same wave vector, manifested by the reappearance of imaginary phonon frequencies. As a result, unlike 1T--MoSH and 2H--MoSeH, the CDW order in 1T--MoSeH cannot be completely suppressed by either strain or carrier doping. 

    \begin{figure}[h!]
    \centering
    \includegraphics[width=8.5cm]{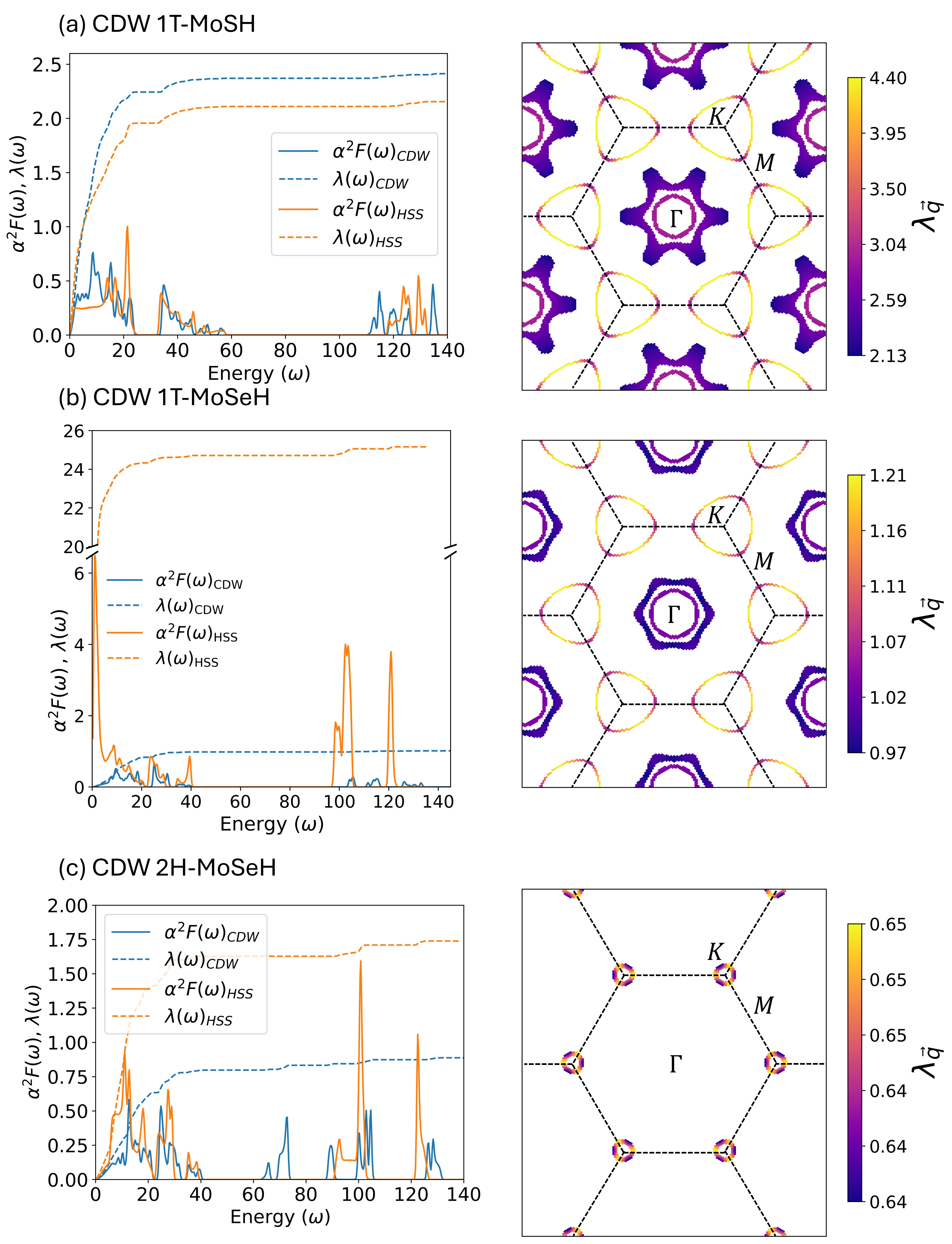}
    \caption{Effect of CDW order on electron--phonon coupling and superconductivity in Janus MoXH monolayers.
    Left panels show the Eliashberg spectral function $\alpha^2F(\omega)$ (solid lines) and the cumulative electron--phonon coupling constant $\lambda(\omega)$ (dashed lines) for the CDW phase (blue) and the corresponding high-symmetry structure (HSS, orange), while right panels display the momentum-resolved electron--phonon coupling strength $\lambda_{\mathbf{q}}$ over the hexagonal Brillouin zone.
    Panels correspond to
    (\textbf{a}) 1T--MoSH,
    (\textbf{b}) 1T--MoSeH, and
    (\textbf{c}) 2H--MoSeH.
    }
    \label{fig:cdw_superconducting}
    \end{figure}

Figure~\ref{fig:cdw_superconducting} summarizes the impact of CDW order on the electron--phonon coupling (EPC) and superconductivity in Janus MoXH monolayers. By comparing the CDW ground states with their corresponding high-symmetry structures stabilized by strain or carrier doping, a clear and material-dependent interplay between CDW formation and superconductivity emerges. For 1T--MoSH, the CDW phase enhances the EPC and superconducting critical temperature. In the CDW ground state, the EPC constant increases relative to the strained HSS reference, accompanied by enhanced low-energy phonon contributions to $\alpha^2F(\omega)$ and a redistribution of the momentum-resolved coupling $\lambda_{\mathbf{q}}$, as shown in Fig.~\ref{fig:cdw_superconducting}(a). This enhancement originates from the stabilization of the soft phonon mode at the $M$ point in the CDW phase, in contrast to the behavior observed in MoSeH-based systems. As a result, CDW-induced lattice distortions do not universally suppress superconductivity; instead, in 1T--MoSH, the CDW phase cooperatively enhances the phonon-mediated pairing interaction.

In sharp contrast, CDW formation suppresses superconductivity in both 1T--MoSeH and 2H--MoSeH by reducing the effective electron--phonon coupling, consistent with previous studies on NbSe$_2$~\cite{cho2018using,kawakami2025dichotomy,zheng2019electron}. In 1T--MoSeH, the CDW phase yields a moderate EPC strength and superconducting transition temperature, whereas the heavily hole-doped HSS exhibits an unrealistically large EPC constant accompanied by a very low characteristic phonon frequency, indicative of an unphysical lattice instability rather than a meaningful superconducting state. In this case, the CDW phase suppresses excessive EPC and stabilizes the lattice at the expense of a reduced $T_c$. A similar suppression of EPC and superconductivity is observed in 2H--MoSeH, where the CDW phase exhibits weaker coupling and a lower $T_c$ compared with the compressively strained HSS. In both MoSeH-based systems, CDW formation quenches strongly localized EPC hotspots present in the high-symmetry structures and redistributes $\lambda_{\mathbf{q}}$ more uniformly across the Brillouin zone, leading to a suppression of superconductivity. AIMD further indicate that the CDW phases remain robust up to temperatures below $T_{\mathrm{CDW}}\sim 50$~K. For all three systems, the superconducting transition temperatures satisfy $T_c < T_{\mathrm{CDW}}$, indicating that superconductivity develops within a pre-established CDW background. 

In summary, we have systematically investigated the origin, stability, and superconducting consequences of charge-density-wave (CDW) order in Janus MoXH ($X=\mathrm{S,Se}$) monolayers by combining first-principles calculations, phonon analysis, electronic susceptibility, and AIMD simulations. We demonstrate that the CDW instability in these systems originates from strong momentum-dependent electron--phonon coupling associated with phonon linewidth enhancement and soft phonon modes at the $M$ point, rather than from Fermi-surface nesting. Real-space structural optimizations using commensurate supercells confirm the formation of a stable $2\times2$ CDW phase, which fully removes the imaginary phonon modes present in the high-symmetry metallic structures. AIMD simulations further show that the CDW phases remain robust up to temperatures below $T_{\mathrm{CDW}}\sim 50$~K, establishing them as genuine ground-state lattice reconstructions. In all systems studied, the superconducting transition temperatures satisfy $T_c < T_{\mathrm{CDW}}$, indicating that superconductivity develops within a pre-existing CDW background.

The interplay between CDW order and superconductivity is found to be strongly material dependent. In 1T--MoSH, CDW formation enhances low-energy phonon contributions to the Eliashberg spectral function, leading to increased electron--phonon coupling and a higher superconducting transition temperature compared with the strained high-symmetry structure. In contrast, for both 1T--MoSeH and 2H--MoSeH, CDW ordering suppresses the electron--phonon coupling and reduces $T_c$, while simultaneously stabilizing the lattice by quenching unphysically large coupling hotspots present in doped or strained high-symmetry configurations. Overall, our results reveal a dichotomous role of CDW order in Janus transition-metal chalcogenide hydrides, acting either cooperatively or competitively with superconductivity depending on the detailed lattice and electronic structure. These findings provide a unified understanding of CDW–superconductivity interplay in low-dimensional Janus materials and suggest viable routes for engineering superconducting properties through controlled manipulation of lattice instabilities.

This work was supported by the Second Century Fund (C2F), Chulalongkorn University. Computational resources were provided by NSTDA, Chulalongkorn University (CU), CUAASC, and NSRF through PMUB projects B05F650021 and B37G660013 (Thailand) (\url{https://www.e-science.in.th}). This work also made use of the ARCHER2 UK National Supercomputing Service (\url{https://www.archer2.ac.uk}) through the UKCP consortium.
    
\bibliography{apssamp}

\end{document}